\documentclass[letterpaper, 10pt, conference]{ieeeconf}
\usepackage[autostyle]{csquotes}
\usepackage{bm}
\usepackage{amssymb}
\usepackage{epsfig}
\usepackage{subfigure}
\usepackage{multicol, blindtext}
\usepackage{amsmath}
\usepackage{MnSymbol}
\usepackage{graphicx}
\usepackage{hyperref}
\usepackage{url}
\usepackage{epstopdf}
\usepackage{float}
\usepackage{amsfonts}
\usepackage{color}
\usepackage{dblfloatfix}
\usepackage{multirow}%
\usepackage{array}
\usepackage{makecell}

\providecommand{\U}[1]{\protect\rule{.1in}{.1in}}
\DeclareMathOperator*{\argmin}{argmin}

\newtheorem{definition}{\rm\textbf{Definition}}
\newtheorem{theorem}{\rm\textbf{Theorem}}

\IEEEoverridecommandlockouts
\title{Reinforcement Learning-based Receding Horizon Control using Adaptive Control Barrier Functions for Safety-Critical Systems$^*$\thanks{$^*$Research partially supported by the NSF under grants ECCS-1931600, DMS-1664644, CNS-1645681, ECCS-2317079, CCF-2200052, CCF-2340776,  CPS-1932162, and IIS-1914792, the DOE under grants DE-AC02-05CH11231 and DE-EE0009696, the ONR under grant N00014-19-1-2571 and the ARPA-E under grant DEAR0001282.}}

\date{August 2022}

\begin{document}
\makeatletter
\newcommand{\linebreakand}{%
  \end{@IEEEauthorhalign}
  \hfill\mbox{}\par
  \mbox{}\hfill\begin{@IEEEauthorhalign}
}

\author{Ehsan Sabouni$^{\dagger,1}$, H.M. Sabbir Ahmad$^{\dagger,1}$, Vittorio Giammarino$^{\dagger,1}$, \\ 
Christos G. Cassandras$^{2}$, Ioannis Ch. Paschalidis$^{3}$, Wenchao Li$^{2}$
\thanks{$^\dagger$These authors contributed equally to this paper.} 
\thanks{$^{1}$Ehsan Sabouni, H.M. Sabbir Ahmad and Vittorio Giammarino are with Division of Systems Engineering, Boston University, Boston, MA 02215, USA {\tt\small \{esabouni, sabbir92, vgiammar\}@bu.edu}}
\thanks{$^{2}$Christos G. Cassandras and Wenchao Li are with the Department of Electrical and Computer Engineering and the Division of Systems Engineering, Boston University, Boston, MA 02215, USA
        {\tt\small \{cgc, wenchao\}@bu.edu}}
\thanks{$^{3}$Ioannis Ch. Paschalidis is with the Department of Electrical and Computer Engineering, Division of Systems Engineering, and Department of Biomedical Engineering, Boston University, Boston, MA 02215, USA
        {\tt\small yannisp@bu.edu}}       
}
\IEEEoverridecommandlockouts
\makeatletter\def\@IEEEpubidpullup{6.5\baselineskip}\makeatother
\maketitle

\begin{abstract}
Optimal control methods provide solutions to safety-critical problems but easily become intractable. Control Barrier Functions (CBFs) have emerged as a popular technique that facilitates their solution by provably guaranteeing safety, through their forward invariance property, at the expense of some performance loss.This approach involves defining a performance objective alongside CBF-based safety constraints that must always be enforced. Unfortunately, both performance and solution feasibility can be significantly impacted by two key factors such as the selection of the cost function and associated parameters, and 
the calibration of parameters within the CBF-based constraints, which capture the trade-off between performance and conservativeness. To address these challenges, we propose a Reinforcement Learning (RL)-based Receding Horizon Control (RHC) approach leveraging Model Predictive Control (MPC) with CBFs (MPC-CBF). In particular, we parameterize our controller and use bilevel optimization, where RL is used to learn the optimal parameters while MPC computes the optimal control input. We validate our method by applying it to the challenging automated merging control problem for Connected and Automated Vehicles (CAVs) at conflicting roadways.
Results demonstrate improved performance and a significant reduction in the number of infeasible cases compared to traditional heuristic approaches used for tuning CBF-based controllers, showcasing the effectiveness of the proposed method. 
\end{abstract}

\section{Introduction} 
Safety-critical systems have applications across a large number of sectors including transportation, industrial automation, smart manufacturing, and medical systems, to name a few. In recent years, CBFs have been extensively applied in safety-critical control using a solution approach based on a sequence of Quadratic Programs (QPs)
\cite{ames2014control}, referred to as QP-CBF. This method is often subject to infeasibility issues due to (i) the lack of any prediction ability making the QP-CBF method myopic, and (ii) inter-sampling control errors due to time discretization. Additionally, CBF constraints require choosing a class $\mathcal{K}$ function that provides a trade-off between safety and performance.  

To address this issue, the authors in \cite{Ahmad_01} proposed an event-triggered control method to tackle primarily the inter-sampling error problem. 
The authors in \cite{Sreenath_01,Sreenath_02,Sreenath_03} proposed MPC-based control using CBFs which can combat the myopic limitations
of QPs through a look-ahead prediction ability. 
However, these works lack systematic means to tune the parameters. Usually, this choice is done heuristically, which can result in sub-optimal controller response and, in the worst case, safety violation. In order to tackle this issue, the authors in \cite{Xiao_02} proposed adaptive CBFs (AdaCBF), to address the conservativeness of CBFs. However, this method becomes challenging to implement in practice because it requires defining \enquote{penalty terms} in the CBF constraints and their corresponding dynamics. Additionally, this method provides no guidance on tuning the CBF constraints to balance the trade-off for optimizing the system response while ensuring safety. 

In the context of AdaCBFs, the authors in \cite{10077790} proposed an end-to-end neural network-based controller design using differential QPs, where the parameters are tuned by means of behavioral cloning \cite{Pomerleau_01}; this performs poorly on out-of-distribution data due to a quadratic compounding of errors over time \cite{Ross_01}. In \cite{Zanon, romero2024actorcritic}, the authors use RL to learn optimal parameters of an MPC controller by using $Q$-learning and actor-critic algorithms. However, their methods require computing gradients by solving KKT conditions and backpropagating through the MPC (which can be nonlinear), respectively, which makes their method computationally expensive and generally intractable.

The QP-CBF method has been applied to safety-critical control problems involving Connected and Automated Vehicles (CAVs) at various conflict points of transportation networks
\cite{9720282, rios2016automated, MAHBUB2020108958, 10421858}.
Aside from the issues addressed above, the solutions to these problems have also been largely limited to the longitudinal dynamics of vehicles, decoupling these from their lateral dynamics. 
In this context, we make the following contributions:
\begin{itemize}
    \item {We propose a parameterized MPC controller with CBFs in order to address the infeasibility issues encountered in the QP-CBF methods. Moreover, we use RL to learn the optimal parameters in the MPC objective function, as well as the parameters in the CBF-based constraints, thus balancing the trade-off between safety and performance.}

    \item {Our proposed approach does not require backpropagating through the MPC-CBF controller, which can be computationally expensive and intractable. As a result, our approach is always computationally efficient.}

    \item {We tackle the problem of CAV control in merging roadways by considering both longitudinal and lateral motions. Besides validating our approach, the problem setting allows us to highlight the generalizability of the learnt controller. We achieve this by training the controller parameters for a single CAV and using them across a set of homogeneous CAVs during deployment.}
\end{itemize}

The paper is organized as follows: in Section~\ref{prelims}, we provide preliminary technical materials. In Section~\ref{sec:problem}, we present the problem formulation. Subsequently, in Section~\ref{NN_formulation}, we present the parameterized MPC-CBF control design along with the details of the RL setting used to learn the controller parameters. This is followed by the details of the multi-agent CAV control problem presented in Section~\ref{CAV_Merging}. Finally, simulation results are presented in Section~\ref{results}, which is followed by conclusions in Section~\ref{conclusion}.
 
\section{Preliminaries} 
\label{prelims}
Consider a control affine system
\begin{align}\label{eq_nonlinear}
    \boldsymbol{\dot{x}}=f(\boldsymbol{x})+g(\boldsymbol{x})\boldsymbol{u},
\end{align}
where $f:\mathbb{R}^n \rightarrow \mathbb{R}^n$ and $g:\mathbb{R}^n \rightarrow \mathbb{R}^{n\times q}$ are locally Lipschitz, $\boldsymbol{x}\in \mathcal{X} \subset \mathbb{R}^n$ denotes the state vector and $\mathbf{u} \in \mathcal{U}\subset \mathbb{R}^q$ the input vector with $\mathcal{U}$ the control input constraint set defined as:
\begin{equation} \label{cntrl_bound}
    \mathcal{U}:=\left\{\boldsymbol{u} \in \mathbb{R}^{q}: \boldsymbol{u}_{\min } \leq \boldsymbol{u} \leq \boldsymbol{u}_{\max }\right\} .
\end{equation}
with $\boldsymbol{u}_{\min},\boldsymbol{u}_{\max} \in \mathbb{R}^q$ and the inequalities are interpreted component-wise . It is assumed that the solution of \eqref{eq_nonlinear} is forward complete.

\subsection{Control Barrier Functions}
\begin{definition}[Class $\mathcal{K}$ function]
    A continuous function $\alpha : [0,a)\rightarrow [0,\infty], a > 0$ is said to belong to class $\mathcal{K}$ if it is strictly increasing and $\alpha(0)=0$.
\end{definition}

\begin{definition}
A set $C$ is forward invariant for system \eqref{eq_nonlinear} if for every $\boldsymbol{x}(0) \in C$, we have $\boldsymbol{x}(t) \in C$, for all $t \geq 0$.
\end{definition}

\begin{definition}[Control barrier function \cite{Ames_01}]
\label{Def_3}
    Given a continuously differentiable function $b:\mathbb{R}^n\rightarrow \mathbb{R}$ and the set $C:=\{\boldsymbol{x} \in \mathbb{R}^n:b(\boldsymbol{x})\geq 0\}$, $b(\boldsymbol{x})$ is a candidate control barrier function (CBF) for the system (\ref{eq_nonlinear}) if there exists a class $\mathcal{K}$ function $\alpha$ such that
    \begin{equation}
        \sup _{\boldsymbol{u} \in \mathcal{U}}\left[L_{f} b(\boldsymbol{x})+L_{g} b(\boldsymbol{x}) \boldsymbol{u}+\alpha(b(\boldsymbol{x}))\right] \geq 0,
        \label{cbf_condition}
    \end{equation}
    for all $\boldsymbol{x} \in C$, where $L_{f}, L_{g}$ denote the Lie derivatives along $f$ and $g$, respectively.
\end{definition}

\begin{definition}[Relative degree]
    The relative degree of a (sufficiently many times) differentiable function $b: \mathbb{R}^n \rightarrow \mathbb{R}$ with respect  to system \eqref{eq_nonlinear} is the number of times it needs to be differentiated along its dynamics until the control $\boldsymbol{u}$ explicitly appears in the corresponding derivative.
\end{definition}

For a constraint $b(\boldsymbol{x}) \geq 0$ with relative degree $m, \ b$ : $\mathbb{R}^{n} \rightarrow \mathbb{R}$, and $\zeta_{0}(\boldsymbol{x}):=b(\boldsymbol{x})$, we define a sequence of functions $\zeta_{i}: \mathbb{R}^{n} \rightarrow \mathbb{R}, i \in\{1, \ldots, m\}$ :
\begin{equation} \label{psi functions}
    \zeta_{i}(\boldsymbol{x}):=\dot{\zeta}_{i-1}(\boldsymbol{x})+\alpha_{i}\left(\zeta_{i-1}(\boldsymbol{x})\right),\; i \in\{1, \ldots, m\},
\end{equation}
where $\alpha_{i}(\cdot),\ i \in\{1, \ldots, m\}$ denotes a $(m-i)^{\text {th }}$ order differentiable class $\mathcal{K}$ function. We further define a sequence of sets $C_i,\ i\in \{1,...,m\}$ associated with \eqref{psi functions} which take the following form,
\begin{equation} \label{C set}
    C_i:=\{\boldsymbol{x} \in \mathbb{R}^n : \zeta_{i-1}(\boldsymbol{x})\geq 0\},\; i \in \{1,...,m\}.
    \end{equation}
\begin{definition}[High Order CBF (HOCBF) \cite{xiao2019HOCBF}]
    Let $C_1,...,C_m$ be defined by \eqref{C set} and $\zeta_1(\boldsymbol{x}),...,\zeta_m(\boldsymbol{x})$ be defined by $\eqref{psi functions}$. A function $b: \mathbb{R}^n \rightarrow \mathbb{R}$ is a High Order Control Barrier Function (HOCBF) of relative degree $m$ for system \eqref{eq_nonlinear} if there exists $(m-i)^{th}$ order differentiable class $\mathcal{K}$ functions $\alpha_i,\ i \in \{1,...,m-1\}$ and a class $\mathcal{K}$ function $\alpha_m$ such that 
    \begin{equation} \label{HOCBF}
        \sup_{\boldsymbol{u}\in \mathcal{U}} [L_f^mb(\boldsymbol{x})+L_gL_f^{m-1}b(\boldsymbol{x})\boldsymbol{u}+S(b(\boldsymbol{x}))+\alpha_m(\zeta_{m-1}(\boldsymbol{x}))] \geq 0
    \end{equation}
    for all $\boldsymbol{x} \in \bigcap_{i=1}^m C_i$. In \eqref{HOCBF}, $L_f^m$ and $L_g$ denotes derivative along $f$ and $g$ $m$ times and one time respectively, and $S(\cdot)$ denotes the remaining Lie derivative along $f$ with degree less than or equal to $m-1$ (ommited for simplicty, see \cite{xiao2019HOCBF}).
\end{definition}

Note that the HOCBF in \eqref{HOCBF} is a general form of the degree one CBF \cite{ames2014control} ($m=1$) and exponential CBF in \cite{nguyen2016exponential}. The following theorem on HOCBFs implies the forward invariance property of the CBFs and the original safety set. The proof is omitted (see \cite{xiao2019HOCBF} for the proof).

\begin{theorem}[\cite{Ames_01}]
\label{cbf_theorem}
    Given a constraint $b(\boldsymbol{x}(t))$ with the associated sets $C_i$'s as defined in (\ref{C set}), any Lipschitz continuous controller $\boldsymbol{u}(t)$, that satisfies (\ref{HOCBF}) $\forall t \geq t_{0}$ renders the sets $C_i$ (including the set corresponding to the actual safety constraint $C_1$) forward invariant for control system \eqref{eq_nonlinear}.
\end{theorem} 

\begin{definition}[Control Lyapunov function (CLF)\cite{ames2012control}] A continuously differentiable function $V:\mathbb{R}^n \rightarrow \mathbb{R}$ is a globally and exponentially stabilizing CLF for \eqref{eq_nonlinear} if there exists constants  $c_i \in \mathbb{R}_{>0}$, $i=1,2$, such that $c_1 ||x||^2 \leq V(x) \leq c_2 ||x||^2$, and the following inequality holds
\begin{equation} \label{CLF}
\inf_{\boldsymbol{u}\in \mathcal{U}} [L_fV(\boldsymbol{x})+L_gV(\boldsymbol{x})u+ \eta(\boldsymbol{x}) ]\leq e,
\end{equation}
where $e$ makes this a soft constraint.
\end{definition}

\subsection{Reinforcement Learning}
\label{reinforcement_learning}
We consider an infinite-horizon discounted Markov Decision Process (MDP) defined by the tuple $(\mathcal{S}, \mathcal{A}, \mathcal{T}, \mathcal{R}, \rho_0, \gamma)$, where $\mathcal{S}$ is the set of states and $\mathcal{A}$ is the set of actions. $\mathcal{T}:\mathcal{S}\times \mathcal{A} \rightarrow \mathcal{P}(\mathcal{S})$ is the transition probability function and $P(\mathcal{S})$ denotes the space of probability distributions over $\mathcal{S}$. The function $\mathcal{R}: \mathcal{S} \times \mathcal{A} \rightarrow \mathbb{R}$ maps state-action pairs to rewards. $\rho_0 \in \mathcal{P}(\mathcal{S})$ is the initial state distribution and $\gamma \in [0,1)$ the discount factor. We model the decision agent as a stationary policy $\pi:\mathcal{S}\rightarrow \mathcal{P}(\mathcal{A})$, where $\pi(a|s)$ is the probability of taking action $a$ in state $s$. The RL objective is to choose a policy that maximizes the expected total discounted reward $J(\pi)=\mathbb{E}_{\tau_{\pi}}[\sum_{k=0}^{\infty}\gamma^k \mathcal{R}(\bm{s}_k,\bm{a}_k)]$, where $\tau_{\pi} = (\bm{s}_0,\bm{a}_0,\bm{s}_1,\bm{a}_1,\dots)$ is a trajectory sampled according to $\bm{s}_0 \sim \rho_0$, $\bm{a}_k\sim\pi(\cdot|\bm{s}_k)$ and $\bm{s}_{k+1}\sim \mathcal{T}(\cdot|\bm{s}_k,\bm{a}_k)$. We also denote the state value function of $\pi$ as $V^{\pi}(\bm{s}) = \mathbb{E}_{\tau_{\pi}}[\sum_{k=0}^{\infty}\gamma^k \mathcal{R}(\bm{s}_k,\bm{a}_k)|\bm{s}_0=s]$ and the state-action value function as $Q^{\pi}(\bm{s},\bm{a}) = \mathbb{E}_{\tau_{\pi}}[\sum_{k=0}^{\infty}\gamma^{k} \mathcal{R}(\bm{s}_k,\bm{a}_k)|\bm{s}_0=\bm{s}, \bm{a}_0=\bm{a}]$. 

\subsection{Model Predictive Control}
The idea of the MPC scheme is at each sampling instance $h = 0,1,...,N-1$, we optimize the predicted future behavior of the system over the finite horizon of time of length $N \geq  2$, and then feed only the first element of the obtained optimal control sequence as a feedback control value for the next sampling interval. In general, the optimization problem includes the dynamics of the system with the enforcement of states and input constraints. In this type of problem, and for reducing the computational complexity, the continuous time differential equation governing the system dynamics is replaced with a discrete-time difference equation.  The formulation of the MPC for the control of a constrained nonlinear time-invariant system governed by the difference equation 
\begin{equation}\label{diff_eqn}
    \bm{x}_{k+1}=f(\bm{x}_k,\bm{u}_k),
\end{equation}
is as follows:
\begin{align} \label{MPC}
    &\min_{\boldsymbol{x}(\cdot),\boldsymbol{u}(\cdot)}  \sum_{h=0}^{N-1} l(\bm{x}_{h|k},\bm{u}_{h|k})+V_N(\bm{x}_{N|k})\\ \nonumber
    \textnormal{subject to} &  \ \   \bm{x}_{h+1|k}=f(\bm{x}_{h|k},\bm{u}_{h|k}),   \ \ \ \ \ h=0,...,N-1, \\ \nonumber
    &\bm{x}_{h|k}  \in \mathcal{X},\bm{u}_{h|k}  \in \mathcal{U},   \ \ \ \ \ \ \ \ \ \ \ \ h=0,...,N-1, \\ \nonumber 
    &\bm{x}_N \in \mathcal{X}_f,
\end{align}
where $k$ is the discrete time index, $h$ is the horizon index, $\boldsymbol{x}_{h|k}$ and $\boldsymbol{u}_{h|k}$ are the state and control input values respectively at step $h$ of the horizon starting at time step $k$. Additionally, $l(\cdot)$ is the stage cost, $V_N$ is the terminal cost, and $\bm{x}_N$ is the terminal state. Finally, $\mathcal{X}_f$ is the set of admissible states for the terminal states, and $\mathcal{X}$ and $\mathcal{U}$ are as defined in Section \ref{prelims}. 
Let $\boldsymbol{u}^*_{0:N-1|k}=\{\boldsymbol{u}^*_{0|k},...,\boldsymbol{u}^*_{N-1|k}\}$ be the solution to the optimization problem in \eqref{MPC} at time step $k$ if it exists. Then, the first element of the solution set will be fed into the system to drive it to the next time step, at which point the control will be re-computed and updated by solving \eqref{MPC}.

\section{Problem Formulation}
\label{sec:problem}

We consider the control of a safety-critical system. We assume that the system has the affine dynamics in \eqref{eq_nonlinear} with control bounds as in \eqref{cntrl_bound}. The system has a certain high-level performance specification associated with it, as reflected in the cost function $l(\boldsymbol{x},\boldsymbol{u})$. In addition to the cost, we have a set of safety constraints $b_j(\boldsymbol{x}) \geq 0, ~j \in \Lambda_S$ (where $b_j$ is continuously differentiable, $\Lambda_S$ is a constraint set). We define the stage cost associated to the system in \eqref{eq_nonlinear} as:
\begin{equation} \label{stage_cost}
   L(\boldsymbol{x}, \boldsymbol{u})=l(\boldsymbol{x}, \boldsymbol{u})+\sum_{j \in \Lambda_S}\mathcal{I}_{\infty}(b_j(\boldsymbol{x})), 
\end{equation}
where we use the indicator function
$$
\mathcal{I}_{\infty}(x)= \begin{cases}\infty, \text { if } x<0 \text { for some } i, \\ 0,  \text{  otherwise. }\end{cases}
$$
to ensure the satisfaction of each $b_j(\boldsymbol{x}) \geq 0$.
Finally, we consider a parameterized state feedback controller $\boldsymbol{u}(\bm{x}|\boldsymbol{\theta})$; our goal is to learn the parameters $\bm{\theta}$ that achieve:
\begin{itemize}
    \item Minimize the expected sum of discounted costs:
    \begin{equation}
        \label{dis_cost}
        \boldsymbol{\theta}^*_{0, \dots, k, \dots} = \argmin_{\boldsymbol{\theta}_{0, \dots, k, \dots}} \mathbb{E}_{\tau_{\boldsymbol{u}}} [\sum_{k=0}^{\infty} \gamma^k L(\boldsymbol{x}_k, \boldsymbol{u}_k (\bm{x}_k|\bm{\theta}_k))], 
    \end{equation}
    where, $\tau_{\boldsymbol{u}}$ are trajectories sampled based on the dynamics in \eqref{diff_eqn} given an initial state distribution at $k = 0$. Here, $\gamma$ is the discount factor which yields a well-posed problem, i.e., the state and state-action value functions, defined in Section~\ref{reinforcement_learning}, are well posed and finite over some sets.

    \item Guarantee the satisfaction of the constraints $b_j(\boldsymbol{x}_k),\ \forall k, ~j \in \Lambda_S$,  and control bounds in \eqref{cntrl_bound}.
\end{itemize}

Given the objective in \eqref{dis_cost}, we will first present the formulation of the parameterized state feedback controller, which in our case is obtained using MPC-CBF, followed by the determination of the parameters in \eqref{dis_cost} learnt using RL.

\section{Parameterized MPC-CBF control Design}
\label{NN_formulation}
In this section, we present the formulation of the parameterized controller $\bm{u}(\bm{x}|\bm{\theta})$ (cf. Section \ref{sec:problem}). Note that we exclude the temporal dependency to streamline the notation. 

\textbf{MPC objective:} We adopt a Receding Horizon Control (RHC) approach using MPC due to its look-ahead prediction ability. The stage cost in \eqref{stage_cost} has two parts: a cost term and a penalty term associated with safety constraints. We encode the safety constraints as hard constraints in the MPC formulation. The cost $l(\boldsymbol{x},\boldsymbol{u})$ in \eqref{stage_cost} can be directly incorporated into the objective function. Note that the cost $l(\boldsymbol{x},\boldsymbol{u})$ is usually selected arbitrarily to reflect some high-level system specification/objective. Consequently, it can be highly nonlinear and/or not differentiable, making it computationally intractable.  Hence, we select a continuous quadratic function of states $\boldsymbol{x}$ and control input $\boldsymbol{u}$ as the stage cost in our MPC objective and parameterize it to optimize \eqref{dis_cost}. We express the parameterized MPC stage cost as follows:
\begin{equation} \label{Obj_param}
 J(\boldsymbol{x}, \boldsymbol{u} | {\boldsymbol{\theta}_o}) = \begin{bmatrix}
\boldsymbol{x}^T & \boldsymbol{u}^T
\end{bmatrix} A(\boldsymbol{\theta}_o) \begin{bmatrix}
\boldsymbol{x}\\ 
\boldsymbol{u}
\end{bmatrix} + B(\boldsymbol{\theta}_o)^T \begin{bmatrix}
\boldsymbol{x}\\ 
\boldsymbol{u}
\end{bmatrix},
\end{equation}
where $\bm{\theta}_o$ is the vector of the learnable parameters of the objective. Without loss of generality, $\bm{\theta}_o$ may consist of weights for state and control components in (\ref{Obj_param})).

\textbf{MPC Constraints:} We deal with the second term in \eqref{stage_cost} by enforcing hard HOCBF constraints in the MPC.
As mentioned in Section \ref{prelims}, HOCBFs, because of their forward invariance property, provide a sufficient condition for the satisfaction of the original constraints. In addition,  
as seen in (\ref{HOCBF}), the transformed HOCBF constraints include class $\mathcal{K}$ functions which can be parameterized to quantify the trade-off between safety and performance. In particular, if the slope of the function is too small, then the controller may be overly conservative resulting in suboptimal performance. Conversely, a steep class $\mathcal{K}$ function in (\ref{HOCBF}) will allow for optimization of the system performance to a greater extent as the set of feasible inputs is larger and trajectories are allowed to approach the boundary of the safety set at a higher rate. However, such a HOCBF constraint \eqref{HOCBF} becomes active only near the unsafe set boundary, requiring a large control input effort which may cause infeasibility due to the control bound in \eqref{cntrl_bound} (see also \cite{10077790}). 
\begin{figure}[t]
\centering
    \includegraphics[scale=0.38]{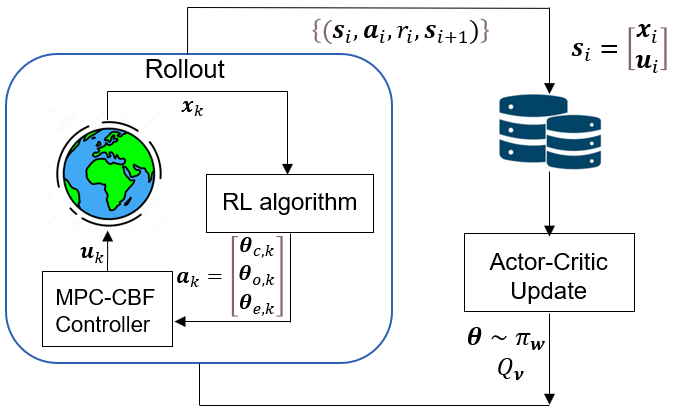}
    \caption{RL training pipeline for parametrized MPC-CBF. The RL agent learns the parameters $[\boldsymbol{\theta}_{c,k} \ \boldsymbol{\theta}_{o,k} \ \boldsymbol{\theta}_{e,k}]^T$ where $\bm{\theta}_o$ is the vector of the learnable parameters of the objective, $\boldsymbol{\theta}_c$ are learnable parameters of the CLF constraint and $\bm{\theta}_e$ is the vector of weights of the penalty terms associated with the relaxation parameters of the CLF constraints. These parameters are then used in the MPC-CBF problem in \eqref{MPC-CBF} which is optimized to compute the optimal control input.}
    \label{RL_pipeline}
\end{figure}
To balance this trade-off, we make the HOCBFs adaptive by parameterizing the class $\mathcal{K}$ function in the HOCBF constraints. Thus, the HOCBF in \eqref{HOCBF} becomes:
\begin{align} \label{HOCBF_param}
    &L_f^m b_j(\boldsymbol{x})+ L_g L_f^{m-1} b_j(\boldsymbol{x}) \boldsymbol{u}+S\left(b_j(\boldsymbol{x}) \mid \boldsymbol{\theta}_c\right) \nonumber \\ &+ \alpha_m\left(\zeta_{m-1}\left(\boldsymbol{x}\right) \mid \boldsymbol{\theta}_c\right) \geq 0, ~~\forall j \in \Lambda_S, 
\end{align}
where $\boldsymbol{\theta}_c$ are the learnable parameters of the HOCBF constraint. For example, a linear parametric class $\mathcal{K}$ function is of the form: $\theta_c^{(1)}b_j(\bm{x})$, where $\theta_c^{(1)} \in \mathbb{R}_{>0}$ is the parameter to be learnt. It is important to note that parameterizing HOCBF constraints does not affect the forward invariance property of the HOCBFs. In order to include penalty terms related to reference/set-point tracking, we define Lyapunov functions and include the corresponding CLF constraints in the MPC, as in prior works (e.g., \cite{Xiao_02}).   Additionally, we parameterize the CLF constraints to also make them adaptive based on the stage cost in \eqref{stage_cost} to optimize \eqref{dis_cost}. A parameterized CLF constraint is expressed as follows:
\begin{equation} \label{CLF_param}
L_fV(\boldsymbol{x})+L_gV(\boldsymbol{x})u+ \eta(\boldsymbol{x} | \boldsymbol{\theta}_c )\leq e,
\end{equation}
where $\boldsymbol{\theta}_c$ are learnable parameters of the CLF constraint, and $e$ is a relaxation parameter for the CLF constraint which is included to make the constraint soft as it is not a safety (critical) constraint.

\textbf{Parameterized MPC-CBF approach}: Finally, the parametrized MPC-CBF problem can be written as follows:
\begin{align}
\begin{split}
    \displaystyle\min_{\boldsymbol{x}_{0:N|k} \atop \boldsymbol{u}_{0:N-1|k}}  & \sum_{h=0}^{N-1} \left(J(\boldsymbol{x}_{h|k},\boldsymbol{u}_{h|k}|\boldsymbol{\theta}_{o,k})  + \bm{\theta}_{e,k}^T \bm{e}_{h|k}^2 \right) \\ 
    &+ V_N(\boldsymbol{x}_{N|k}|\boldsymbol{\theta}_{o,k}), \\
    \label{MPC-CBF}
\end{split} \\
     \textnormal{subject to} &  \ \   \boldsymbol{x}_{h+1|k}=f(\boldsymbol{x}_{h|k},\boldsymbol{u}_{h|k}),    \ \ \ \ \ h=0,\dots,N-1, \nonumber \\ 
    &\eqref{cntrl_bound}, \eqref{HOCBF_param}, \eqref{CLF_param}, \ \ \ \ \ \ \ \ \ \ \ \ \ \ \ \ \ \ h=0,\ldots,N-1, \nonumber \\ 
    & \boldsymbol{x}_{h|k} \in \mathcal{X}, x_N \in \mathcal{X}_f. \nonumber
\end{align}
where $\bm{\theta}_e$ is the vector of weights of the penalty terms associated with the relaxation parameters of the CLF constraints. The CBF (or HOCBF) constraints correspond to the safety constraints $b_j(\boldsymbol{x})$, $\forall j \in \Lambda_S$.

\textbf{RL algorithm}: We denote the learnable parameters of the MPC-CBF as $\boldsymbol{\theta}_k = [\boldsymbol{\theta}_{c,k} \ \boldsymbol{\theta}_{o,k} \ \boldsymbol{\theta}_{e,k}]^T$, where $\boldsymbol{\theta}_k \in \mathbb{R}^{|\boldsymbol{\theta}|}_{>0}$. To learn the optimal parameters corresponding to \eqref{dis_cost}, we set the state space $\mathcal{S} = \mathcal{X} \times \mathcal{U}$ as defined in Section \ref{prelims} and the action space $\mathcal{A} = \mathbb{R}^{|\boldsymbol{\theta}|}_{>0}$. We define the reward $\mathcal{R}(\bm{s}_k) = -L(\bm{s}_k)$, with $L(\bm{s}_k)$ in \eqref{stage_cost}. Finally, we use RL to learn the optimal parameters $\bm{\theta}_k(\bm{s}_k)$ which change at every time step $k$ as a function of the state $\bm{s}_k$. Specifically, we parameterize the optimal controller parameter policy using a neural network with parameters $\boldsymbol{w}$ and train it in order to maximize the expected discounted reward $\mathcal{R}(\bm{s}_k)$ (i.e., solve \eqref{dis_cost}). This step can be implemented using any actor-critic RL algorithms \cite{lillicrap2015continuous, haarnoja2018soft}. The optimal parameters $\bm{\theta}_k$ are then sampled from the learnt policy, i.e., $\boldsymbol{\theta}_k \sim  \pi_{\bm{w}}(\bm{s}_k)$. The overall pipeline of RL with MPC-CBF is shown in Fig. \ref{RL_pipeline}. Note that modern Deep RL actor-critic algorithms parameterize both the actor $\pi_{\bm{w}}$ and the critic $Q_{\bm{\nu}}$ ($\bm{\nu}$ is a vector of the parameters of the critic network, which can be implemented using a neural network). The critic is updated regressing the Bellman equation as in \cite{mnih2013playing} and the actor is then updated by means of policy gradient which exploits the reparameterization trick \cite{lillicrap2015continuous, haarnoja2018soft}.

\section{Multi-Agent Control of CAVs}
\label{CAV_Merging}

In this section, we review the setting for CAVs whose motions are coordinated at conflict areas of a traffic network. This includes merging roads, signal-free intersections, roundabouts, and highway segments where lane change maneuvers take place. We define a Control Zone (CZ) to be an area within which CAVs can communicate with each other or with a coordinator (e.g., a Road-Side Unit (RSU)) which is responsible for facilitating the exchange of information (but not control individual vehicles) inside this CZ. As an example, Fig. \ref{fig:merging} shows a conflict area due to vehicles merging from two single-lane roads and a single Merging Point (MP) that the vehicles must cross from either road \cite{xiao2021decentralized}.
More generally, the CZ may include a set of MPs that each CAV has to cross; for instance in a 4-way intersection with two lanes per direction there are 32 total MPs \cite{9720282}.
\begin{figure}[b]
\centering
$\hspace{-4mm}$\includegraphics[scale=0.4]{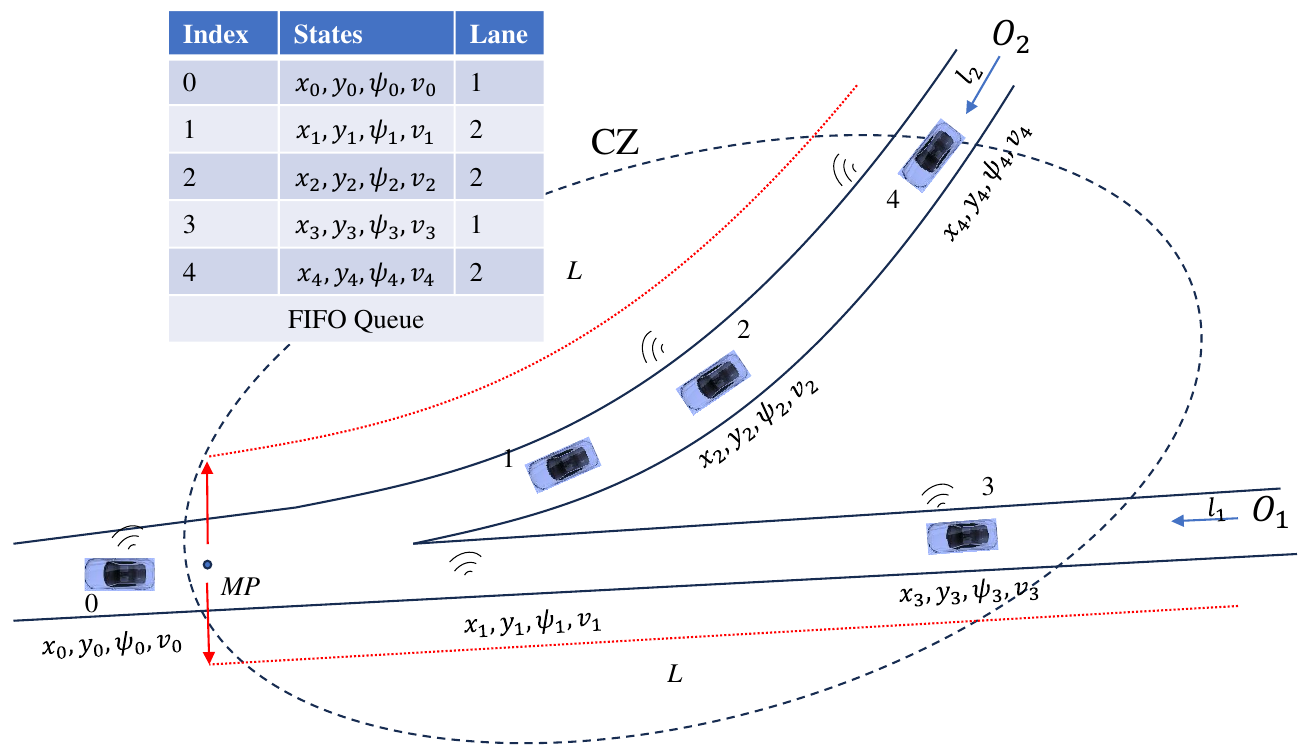} \caption{The merging control problem for CAVs}%
\label{fig:merging}%
\end{figure}
Let $F(t)$ be the set of indices of all CAVs located in the
CZ at time $t$. A CAV enters the CZ at one of several origins (e.g., $O_1$ and $O_2$ in Fig. \ref{fig:merging}) and leaves at one of possibly several exit points (e.g., $MP$ in Fig. \ref{fig:merging}). 
The index $0$ is used to denote a CAV that has just left the CZ. Let $S(t)$ be the
cardinality of $F(t)$. Thus, if a CAV arrives at time $t$, it is assigned the
index $S(t)+1$. All CAVs indices in $F(t)$ decrease by one when a CAV passes over
the MP and the vehicle whose index is $-1$ is dropped.

The vehicle dynamics for each CAV $i\in F(t)$ along its lane in a given CZ are assumed to be of the form
\begin{equation} \label{VehicleDynamics}
\underbrace{\left[\begin{array}{c}
\dot{x}_i \\
\dot{y}_i \\
\dot{\psi}_i \\
\dot{v}_i
\end{array}\right]}_{\dot{\boldsymbol{x}}_i(t)}=\underbrace{\left[\begin{array}{c}
v_i \cos \psi_i \\
v_i \sin \psi_i \\
0 \\
0
\end{array}\right]}_{f\left(\boldsymbol{x}_i(t)\right)}+\underbrace{\left[\begin{array}{cc}
0 & 0 \\
0 & 0 \\
0 & v_i / (l_f+l_r) \\
1 & 0
\end{array}\right]}_{g\left(\boldsymbol{x}_i(t)\right)} \underbrace{\left[\begin{array}{c}
u_i \\
\phi_i
\end{array}\right]}_{\boldsymbol{u}_i(t)},
\end{equation}
where $x_i(t), y_i(t), \psi_i(t), v_i(t)$ represent the current longitudinal position, lateral position, heading angle, and speed, respectively. $u_i(t)$ and $\phi_i(t)$ are the acceleration and steering angle (controls) of vehicle $i$ at time $t$, respectively, $g\left(x_i(t)\right)=$ $\left[g_u\left(\boldsymbol{x}_i(t)\right), g_\phi\left(\boldsymbol{x}_i(t)\right)\right]$. 
There are two objectives for each CAV, as detailed next. \\
 {\bf Objective 1} (Minimize travel time): Let $t_{i}^{0}$ and $t_{i}^{f}$
denote the time that CAV $i\in F(t)$ arrives at its origin
and leaves the CZ at its exit point, respectively. We wish to minimize the travel time
$t_{i}^{f}-t_{i}^{0}$ for CAV $i$.\\
{\bf Objective 2} (Minimize energy consumption): We also wish to minimize
the energy consumption for each CAV $i$:
\begin{equation}
J_{i}(u_{i}(t),\phi_i(t),t_{i}^{f})=\int_{t_{i}^{0}}^{t_{i}^{f}}\mathcal{C}_i(|u_{i}(t)|)dt,
\end{equation}
where $\mathcal{C}_i(\cdot)$ is a strictly increasing function of its argument.

A comfort objective may also be included (e.g., when the CZ includes curved road segments subject to centrifugal forces, but we omit it here for simplicity). We consider next the following constraints.

\textbf{Objective 3} (Center lane following): We wish to minimize the distance of CAV $i$ from the center lane along the path. Let $d_c(x_i(t),y_i(t))$ be the metric representing the distance of the car from the center of the lane at time $t$. Then, the objective can be expressed as follows:
\begin{equation} \label{centr_lane_obj3}
J_{i}(u_{i}(t),\phi_i(t),t_{i}^{f})=\int_{t_{i}^{0}}^{t_{i}^{f}}d_c(x_{i}(t),y_{i}(t))^2dt,
\end{equation}

{\bf Constraint 1} (Safety constraint): Let $i_{p}$ denote the index of
the CAV which physically immediately precedes $i$ in the CZ (if one is
present). We require that the distance between vehicles $i$ and $i_p$ remains safe by defining a speed dependent ellipsoidal safe region $b_{1}\left(\boldsymbol{x}_i, \boldsymbol{x}_{i_p}\right)$ as follows:
\begin{equation}\label{safety}
    b_{1}\left(\boldsymbol{x}_i, \boldsymbol{x}_{i_p}\right):=\frac{\left(x_i(t)-x_{i_p}(t)\right)^2}{\left(a_i v_i(t)\right)^2}+\frac{\left(y_i(t)-y_{i_p}(t)\right)^2}{\left(b_i v_i(t)\right)^2}-1 \geq 0,
\end{equation}
where $a_i$, $b_i$ are weights adjusting
the length of the major and minor axes of the ellipse as shown in Fig. \ref{fig:ellipsoid} and the size of the safe region depends on the speed of CAV $i$.

\begin{figure}[h]
\centering
\includegraphics[scale=0.43]{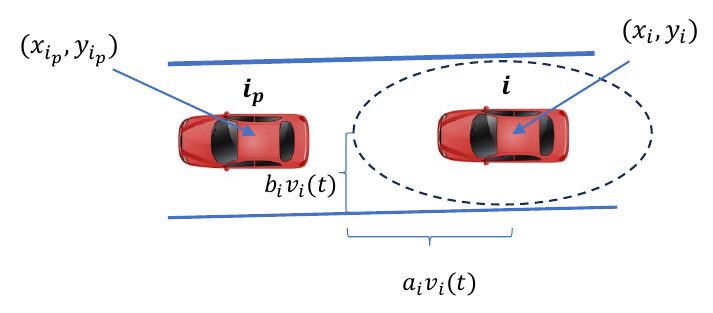} \caption{The ellipsoid for safety}%
\label{fig:ellipsoid}%
\end{figure}

{\bf Constraint 2} (Safe merging): Each CAV $i$ has to maintain safe distancing relative to the CAV that merges immediately ahead of it, denoted by $i_c$. This is ensured by enforcing a constraint on CAV $i$ denoted by $b_2\left(\boldsymbol{x}_i, \boldsymbol{x}_{i_c}\right)$ at the merging point $m$ at time $t_i^m$ when it crosses the MP. $b_2\left(\boldsymbol{x}_i, \boldsymbol{x}_{i_c}\right)$ is defined as follows:
\begin{equation}
\label{SafeMerging}
b_{2}=r_i(t) \mu_i(t)-r_{i_c}(t) \mu_{i_c}(t)-\Phi\left(x_i(t), y_i(t)\right) v_i(t)-\delta \geq 0,
\end{equation}
where, $r_{i} =\sqrt{x_{i}^2+y_{i}^2}, \ \mu_{i}=\tan ^{-1} \left(\frac{y_{i}}{x_{i}}\right)$; similarly, $r_{i_c} =\sqrt{x_{i_c}^2+y_{i_c}^2}, \mu_{i_c}=\tan ^{-1} \left(\frac{y_{i_c}}{x_{i_c}}\right)$. Here $\Phi : \mathbb{R}^2 \rightarrow \mathbb{R}$ may be any continuously differentiable function as long as it is strictly increasing and satisfies the boundary conditions $\Phi(x_i(t_i^0),y_i(t_i^0) )=0$ and $\Phi(x_i(t_i^{m}),y_i(t_i^{m}))=\varphi$.
Here $\varphi$ denotes the reaction time (as a rule, $\varphi=1.8s$ is used,
e.g., \cite{Vogel2003}) and $\delta$ is a given minimum safe distance, (see \cite{10422265} for details). 

The determination of CAV $i_c$ depends on the policy adopted for sequencing CAVs through the CZ, such as First-In-First-Out (FIFO) based on the arrival times of CAVs, Shortest Distance First (SDF) based on the distance to the MP or any other desired policy. It is worth noting that this constraint only applies at a certain time $t_{i}^{m}$ which obviously depends on how the CAVs are controlled. As an example, in Fig. \ref{fig:merging} under FIFO, we have $j=i-1$ and $t_i^m=t_i^f$ since the MP defines the exit from the CZ.

{\bf Constraint 3} (Vehicle limitations): There are constraints
on the speed and control inputs for each $i\in F(t)$:
\begin{align}\label{VehicleConstraints1}
v_{min} \leq v_i(t)\leq v_{max}, \forall t\in[t_i^0,t_i^m], \\
u_{min}\leq u_i(t)\leq u_{max}, \forall t\in[t_i^0,t_i^m], \nonumber \\
 \phi_{min}\leq \phi_i(t)\leq \phi_{max}, \forall t\in[t_i^0,t_i^m], \label{VehicleConstraints2}
\end{align}
where $v_{max}> 0$ and $v_{min} \geq 0$ denote the maximum and minimum speed allowed
in the CZ for CAV $i$, $u_{min}<0$ and $u_{max}>0$ denote the minimum and maximum
acceleration for CAV $i$, respectively, and, $\phi_{min}$ and $\phi_{max}$ correspond to the minimum and maximum steering respectively. 

{\bf Constraint 4} (Road boundaries): To ensure the vehicle stays within the road boundaries we impose two constraints defined as $b_{rc}$ and $b_{lc}$ for the right and left boundary of the road respectively, as follows:
\begin{equation}\label{road_boundary_1}
     b_{3}=\left(x_i(t)-x_{i,l c}(t)\right)^2+\left(y_i(t)-y_{i,l c}(t)\right)^2-r_{l c}^2 \geq 0,
\end{equation}
\begin{equation}\label{road_boundary_2}
b_{4}=\left(x_i(t)-x_{i,r c}(t)\right)^2+\left(y_i(t)-y_{i,r c}(t)\right)^2-r_{r c}^2 \leq 0  , 
\end{equation}
where $x_{i,lc}(t)$ and $x_{i,rc}(t)$ are the longitudinal coordinates and $y_{i,lc}(t)$ and $y_{i,rc}(t)$ are the lateral coordinates of the left and right boundary respectively for vehicle $i$ at time $t$ and $r_{rc} \in \mathbb{R}_{>0}$, $r_{lc} \in \mathbb{R}_{>0}$ are constants. 

\textbf{Derivation of HOCBF Constraints:} We limit ourselves to the derivation of the HOCBF constraint for only one of the constraints, namely \eqref{road_boundary_1}. All other CBF (or HOCBF) constraints can be similarly derived. First, we define $\zeta_0(\bm{x}) = b_3(\bm{x})$ as shown in Section \ref{prelims}. Next, we determine the parameterized CBF constraint by defining $\zeta_1(\bm{x}|\bm{\theta}_{c,k})$ as follows:
\begin{align}\label{CBF_example_1}
        \zeta_1(\bm{x}_i|\bm{\theta}_{c,k}) =  L_fb_3(\bm{x}_i) 
    + \alpha_1(b_3(\bm{x}_i)|\bm{\theta}_{c,k}) \ge 0,
\end{align}
where $L_fb_3(\bm{x}_i)$ can be obtained as follows:
\begin{align*}
    \dot{b}_3(\bm{x}_i) =  \underbrace{2(\left(x_i-x_{i,l c}\right)v_i\textrm{cos}\psi_i+ \left(y_i-y_{i,l c}\right)v_i\textrm{sin}\psi_i)}_{L_fb_3(\bm{x}_i)} + 0\bm{u}_i. \nonumber 
\end{align*}
where $\alpha_1$ is a parameterized class $\mathcal{K}$ function. Observe that the control inputs do not appear in the first order CBF constraint in \eqref{CBF_example_1}. Therefore, we define a parameterized HOCBF constraint by deriving $\zeta_2(\bm{x}_i|\bm{\theta}_{c,k})$, as in (\ref{HOCBF}):
\begin{align} \label{HOCBF_example}
    \zeta_2(\bm{x}_i|\bm{\theta}_{c,k}) =&  L_f^2b_3(\bm{x}_i)+L_gL_fb_3(\bm{x}_i)\bm{u}_i+ S(b_3(\bm{x}_i)|\bm{\theta}_{c,k})+ \nonumber\\
    &\alpha_2(\zeta_1(\bm{x}_i)|\bm{\theta}_{c,k}) \ge 0,
\end{align}
where each term can be calculated as follows:
\begin{align*}
    \dot{\zeta}_1(\bm{x}_i|\bm{\theta}_{c,k}) =&  \underbrace{\frac{\partial\dot{b}_3}{\partial\bm{x}_i} f(\bm{x}_i)}_{L_f^2b_3(\bm{x}_i)} + \underbrace{\frac{\partial\dot{b}_3}{\partial\bm{x}_i} g(\bm{x}_i)}_{L_gL_fb_3(\bm{x}_i)}\bm{u_i}+ \underbrace{\dot{\alpha}_1(b_3(\bm{x}_i)|\bm{\theta}_{c,k})}_{S(b_3(\bm{x}_i)|\bm{\theta}_{c,k})}\nonumber 
\end{align*}
where $\frac{\partial\dot{b}_3}{\partial\bm{x}_i} = [2v_i\text{cos}\psi_i \ 2v_i\text{sin}\psi_i \ -2(x_i-x_{i,l c})v_i\text{sin}\psi_i + 2(y_i-y_{i,l c})v_i\text{cos}\psi_i \ 2\left(x_i-x_{i,l c}\right)\textrm{cos}\psi_i+ 2\left(y_i-y_{i,l c}\right)\textrm{sin}\psi_i ]^T
$. The relative degree of this HOCBF is 2 as the control input appears with the second derivative. The parametarized HOCBF \eqref{HOCBF_example} constraint is included in the parameterized MPC-CBF controller.
The minimization MPC-CBF problem defined in \eqref{MPC-CBF} can  now be formulated. It is worth mentioning that the travel time minimization Objective 1 can be indirectly achieved by defining a CLF constraint of the form in \eqref{CLF} with $V = (v_i-v_{des})^2$ where $v_{des}$ can be set arbitrarily for CAV $i$; thus, speed is maximized to achieve time minimization. As a result, the optimization problem in \eqref{MPC-CBF} is expressed with the objective function $J$ defined in the form of \eqref{Obj_param} subject to the CBF constraints of the actual constraints in  \eqref{safety}, \eqref{SafeMerging}, \eqref{VehicleConstraints1}, \eqref{road_boundary_1}, \eqref{road_boundary_2} and the control bounds in \eqref{VehicleConstraints2}. This problem is then solved numerically as  detailed in the next section.

\section{SIMULATION RESULTS}

\begin{figure}[b]
\centering 
\includegraphics[scale=0.25]{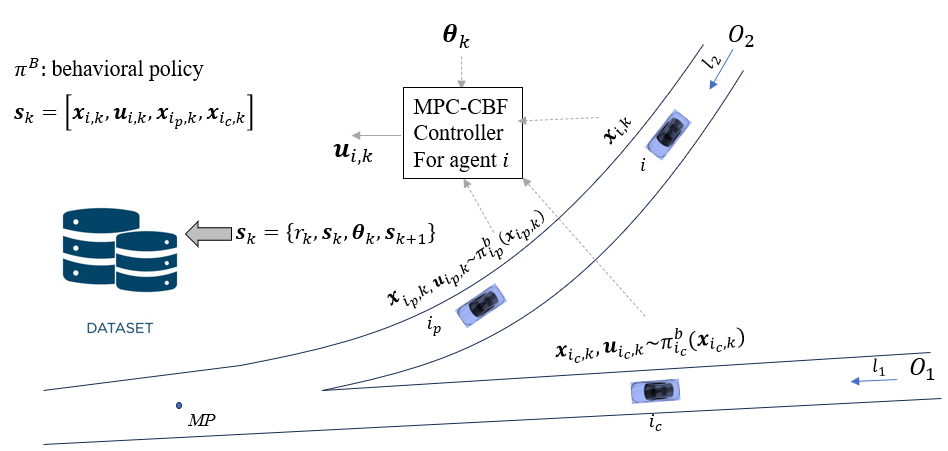} \caption{Illustration of the scenario used to generate rollouts during RL training.}
\label{training_illustration}
\end{figure}

\label{results}
\begin{figure*}[th]
\centering
\includegraphics[scale=0.55]{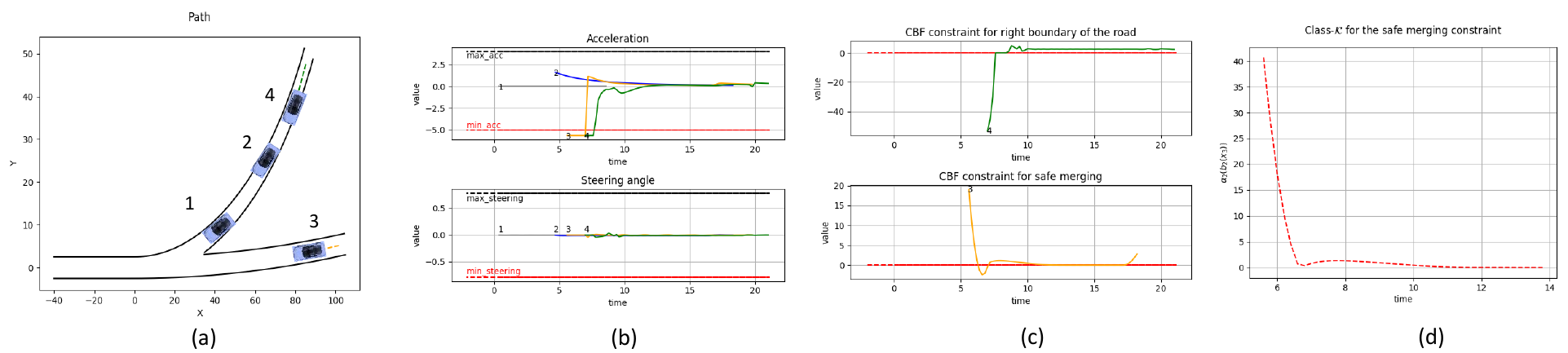}
\caption{Simulation results of the scenario depicted in Fig \ref{fig:merging} with the baseline approach. \textbf{(a)}: A screenshot of the simulation at a point where vehicles $3$ and $4$ encounter infeasibility, as indicated by yellow and green dashes. \textbf{(b)}: Steering angle and acceleration profiles of all vehicles, showing clearly that vehicles $3$ and $4$ violated the bound in their control inputs. \textbf{(c)}: CBF constraint values of right boundary of the road and safe merging constraints of vehicles $4$ and $3$, respectively. As both plots are well below zero, there is an obvious violation. \textbf{(d)}: Evolution of the class $\mathcal{K}$ function $\alpha_2(b_2(\boldsymbol{x}))$ values for the safe merging constraint of CAV 3.}
\label{fig:scenario1}%
\end{figure*}
\begin{figure*}[th]
\centering
\includegraphics[scale=0.55]{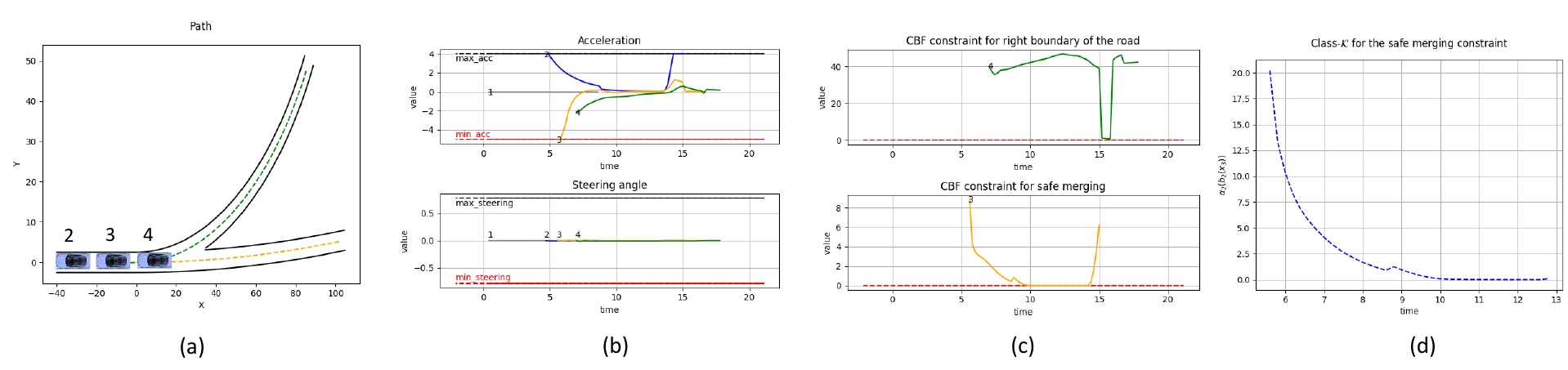} 
\caption{Simulation results of the scenario depicted in Fig \ref{fig:merging} with our proposed approach. \textbf{(a)}: A screenshot of the simulation at the point where vehicles $3$ and $4$ successfully finished their paths without encountering any infeasiblity, indicated by yellow and green dashes. \textbf{(b)}: All vehicles acceleration and steering angles profiles are within the bounds. \textbf{(c)}: CBF constraint values of right boundary of the road and safe merging constraints of vehicles $4$ and $3$, respectively. In contrast to Fig \ref{fig:scenario1}, the values in this plot are all above zero. \textbf{(d)}: Evolution of the class $\mathcal{K}$ function $\alpha_2(b_2(\boldsymbol{x}))$ values for the safe merging constraint of CAV 3.}
\label{fig:scenario2}%
\end{figure*}


\begin{table*} 
        \caption{CAVs metric comparison under the proposed framework and the baseline. The mean and standard deviation were calculated from 10 simulations with different random initial states and arrival times}
        \centering
        \begin{tabular}{|c|c|c|c|c|c|}
            \cline{1-6}
            Item & \multicolumn{4}{c|}{MPC-CBF  w/o RL} & MPC-CBF w/ RL \\
        \hline
          & Conservative & Moderately conservative & Moderately aggressive & Aggressive  & - \\
        \hline
         Ave. travel time & 11.98 $\pm 0.1$ & 11.87 $\pm 0.11$ & 11.82 $\pm 0.1$ &11.78 $\pm 0.1$  & $\boldsymbol{10.94}$ $\pm$ 0.11\\
        \cline{1-6}
        Ave. $\frac{1}{2} u^2$ &  12.19 $\pm 1.23$ & 11.68 $\pm 1.29 $ & 12.55$\pm 1.62$ & 12.83 $\pm 1.75$ & $\boldsymbol{8.70}$ $ \pm$ 1.16\\
        \cline{1-6}
        Ave. fuel consumption &  10.52 $\pm 0.33$ & 10.84 $\pm 0.29$ & 11.35 $\pm 0.3$ & 11.14 $\pm 0.31$  & $\boldsymbol{7.20}$ $\pm 0.27$\\
        \cline{1-6}
        Total infeasibility & 259 & 185 & 199 & 209 & $\boldsymbol{71}$ \\
        \hline
        \end{tabular}
        \label{Table event}
\end{table*}

All algorithms in this paper were implemented using \textsc{Python}. We used the CasADi solver (cf. \cite{Andersson2018}) to solve the minimization problem formulated in \eqref{MPC-CBF}, and \textsc{solve-ivp} was used to solve for the vehicle dynamics \eqref{VehicleDynamics}. In the following, we discuss the MPC-CBF controller and RL setting details as well as the hyperparameters used in the simulation, followed by numerical results. In order to guarantee reproducibility, our code is provided here: \href{https://github.com/EhsanSabouni/CDC2024_RL_adpative_MPC_CBF/tree/main}{\textit{\underline{link}}}\footnote{github.com/EhsanSabouni/CDC2024\_RL\_adpative\_MPC\_CBF/tree/main}.

\textbf{MPC-CBF and model parameters}: 
We have considered the merging problem shown in Fig. \ref{fig:merging} where CAVs arrive at the predefined CZ according to Poisson arrival processes with a given arrival rate. The initial speed $v_{i}(t_{i}^{0})$ is also randomly generated with a uniform distribution over $[5 \textnormal{m/s}, 15\textnormal{m/s}]$ at the origins $O_1$ and $O_2$, respectively. The
hyper parameters 
for solving the problem \eqref{MPC-CBF} are: $L = 100\textnormal{m}, \varphi = 1.2\textnormal{s}, \delta = 3.74\textnormal{m}, u_{max} = 4 \ \textnormal{m/s}^2, u_{min} = -5, \phi_{min} = -\pi/4, \phi_{max} = \pi/4, \ \textnormal{m/s}^2, v_{max} = 20\textnormal{m/s}, v_{min} = 0\textnormal{m/s},v_{des} = 15\textnormal{m/s}, N = 5$. The sampling rate of the discretization and the control update period for control is $\Delta =0.2$s.

In our simulations, we included the computation of a more realistic energy consumption model \cite{kamal2012model} to supplement the simple surrogate $ \ell ^2$-norm ($u^2$) model in our analysis:
\begin{equation*}
f_v(t)=f_{\textrm{cruise}}(t)+f_{\textrm{accel}}(t),
\end{equation*}
\begin{equation*}
 f_{\textrm{cruise}}(t) = \omega_0+\omega_1v_i(t)+\omega_2v^2_i(t)+\omega_3v^3_i(t),
\end{equation*}
\begin{equation*}
    f_{\textrm{accel}}(t) =(r_0+r_1v_i(t)+r_2v^2_i(t))u_i(t),
\end{equation*}
where we used typical values for parameters $\omega_1,\omega_2,\omega_3,r_0,r_1$ and, $r_2$ as reported in \cite{kamal2012model}.

\textbf{RL algorithm and hyperparameters}: 
 The input for the actor neural network is 14 observations (i.e., CAV $i$ states, control inputs, CAV $i_p$ states, CAV $i_c$ states). Both actor and critic neural networks have two hidden layers with 512 units per layer. Actor network dimensions depend on two factors: the number of terms included in the objective functions defined in \eqref{Obj_param} and the type of class $\mathcal{K}$ functions used in such constraints (i.e., linear, exponential, polynomial and so on). Although the problem discussed in Sec.~\ref{CAV_Merging} is a multi-agent problem, we learn the decentralized controller for a single controller assuming the agents are homogeneous. Each CAV/agent at most interacts with two other CAVs ($i_p$ and $i_c$) given in \eqref{safety}, \eqref{SafeMerging}. Therefore, we train an agent $i$ in the presence of two other agents ($i_p$ and $i_c$) and sample their control input from two behavior policies $\pi^{B}_{i_c}$ and $\pi^B_{i_p}$ respectively to ensure sufficient exploration and generalization across all CAVs. Fig.~\ref{training_illustration} illustrates the scenario used to generate the rollouts during training. 
 
 Furthermore, this allows us to define a reward function that depends only on CAV $i$, rather than on the total number of CAVs. The choice of a reward function in the CAV settings can include a number of metrics, including travel time, $\mathrm{L}^2$-norm of control inputs, fuel consumption, and passenger discomfort of the CAV $i$. Since we aim to minimize all of these metrics, the reward can be defined as a negative cost by combining all the metrics above as follows:
\begin{align} \label{SAC_reward}
\mathcal{R}_i=  & -L_i = -( \beta_1||u_i||_2^2 + \beta_2 ||\phi_i||^2 +\nonumber \\ & \beta_3 ||v_i - v_{des}||_2^2 + \beta_4 ||\psi_i - \psi_{des}||^2_2 + \beta_5 f_v + \mathcal{I}_{\infty}),
\end{align}
where $\beta_q$ $\in [0, 1],$ $\forall q $ and $\mathcal{I}_{\infty}$, denotes the infeasiblity penalty 
which is set to be much higher than other terms in the reward function. The actor network output is fed into the optimization problem formulated in \eqref{MPC-CBF} which then outputs the control inputs for the agents.

We train our agent using a soft actor-critic \cite{haarnoja2018soft} for 300000 steps with the use of GPU NVIDIA RTX A5000. The hyperparameters for the implemented algorithm are as follows: learning rate $l_r = 10^{-5}$ and $l_r = 10^{-4}$ for actor and critic, respectively. 
In implementation, the actor network inputs are normalized between the range $[-1,1]$.

\textbf{Numerical Results}: We present simulation results for $50$ random initialized vehicles with two sets of parameters $\boldsymbol{\theta}_k$: the first is our proposed method with a time-varying class $\mathcal{K}$ function and objective weights, while the second is a fixed class $\mathcal{K}$ function (linear function) and fixed objective weights that can serve as a baseline. The reason for this choice for the baseline stems from the fact that the linear class $\mathcal{K}$ functions are very common in the literature as they are easy to tune heuristically. To ensure we pick both conservative and aggressive parameter values in the baseline, the whole parameter range is divided into four sets (i.e., conservative to aggressive). To see the exact values and possibly run the experiments, please refer to the link of the simulation code provided earlier. This allows us to test the effect of fixed linear class $\mathcal{K}$ functions across all parameter spaces. 

The performance of both methods in terms of average travel time, $ \ell ^2$-norm of the acceleration, average fuel consumption and total number of infeasible cases (i.e., the number of times there is no solution for the minimization problem formulated in \eqref{MPC-CBF}) are compared and illustrated in Table \ref{Table event}. According to the results, the proposed method not only improves the desired metrics defined in the reward function (\ref{SAC_reward}), but also reduces the infeasible instances by almost 65$\%$. It is worth noting that an ``infeasible'' case \emph{does not necessarily imply a constraint violation}, since violating a CBF constraint does not always imply the violation of an original constraint as CBF constraints are only sufficient, but not necessary, conditions.

To gain some insight into these improvements, we have created a scenario that can be considered a proof of concept. A similar scenario as depicted in Fig \ref{fig:merging} is used in two experiments, one in which we assign fixed values to the set of parameters, $\boldsymbol{\theta}_k$, and another using our proposed method. The results of these two experiments are illustrated in Figures \ref{fig:scenario1} and \ref{fig:scenario2}, respectively. As it can be seen from Fig. \ref{fig:scenario1}a, the CAV 4 trajectory encounters an infeasiblity (i.e., the obtained control input violated its bounds as shown in Fig. \ref{fig:scenario1}b). This infeasiblity happens due to the conservativeness of the fixed class $\mathcal{K}$ function which then violates the HOCBF constraints defined in \eqref{road_boundary_2} as shown in Fig. \ref{fig:scenario1}c. Similarly, CAV 3 encounters an infeasiblity  and fails to satisfy CBF constraint of safe merging defined in \eqref{SafeMerging} as shown in Fig. \ref{fig:scenario1}c.
However, with our proposed method, both CAV 3 and CAV 4 are able to finish their path completely shown in Fig. \ref{fig:scenario2}a and obtain control inputs within the defined bounds satisfying the HOCBF constraints. 

\section{CONCLUSION}
\label{conclusion}
We have proposed a control method based on RHC using MPC with CBFs which can provably guarantee safety in safety-critical control systems. In order to tackle the issues of performance as well as feasibility, we use RL to learn the parameters of our MPC-CBF controller by optimizing system performance encapsulated in properly defined rewards/costs. We have validated our approach on a multi-agent CAV merging control problem at conflicting roadways. As the proposed approach involves a single-agent, we use RL to train the controller for one CAV and use it across other agents assuming they are homogeneous to evaluate the generalizeability of the proposed approach. Experimental results show that our proposed approach performs better compared to the baseline controller with heuristically selected parameters and also generalizes to unseen scenarios. Future directions will include extending this framework to multi-agent settings where homogeneity assumption does not hold, such as mixed traffic, where CAVs coexist with HDVs.

\end{document}